\documentclass{article}
\usepackage{amsmath}
\title{ Matter Field, Dark Matter and Dark Energy }
\author{Masayasu Tsuge}
\date{ Hiruandon Laboratory\\Tawaradai,Sijonawate,Osaka 575-0013,Japan}
\begin{document}
\maketitle
\begin{abstract}
\quad The electroweak theory indicates the possibility of an unknown space-time where mass and the electromagnetic field do not exist.
As this new spase-time, we propose the quaternion space-time where is one-dimensional real space-time and three dimensional imaginary space-time.
In real space-time part, the lepton field exists and in imaginary space-time part, the quark field exists.
And the bosonic state formed from spinor-anti-spinor exist in real space-time part.
All spinors are massless and two component Weyl spinor.
We define the Weyl spinor phase as dark energy.
If the origin of quark is three dimensional imaginary space-time, we can  understand that all quark fields are confined and cannot take out and that fermions are three generations and that quarks are found in mixing states.
Moreover, because the kinetic energy of quark in three dimensional imaginary space-time is presumed to be a negative quantity, quark will contribute negatively to pressure in dark energy.
When the electromagnetic field $A_{\mu}(x)$ is generated, two Weyl spinors becomes a massive four component spinor due to the action of the energy flow.
Furthermore,various problems regarding the early universe are discussed based on our mass creation mechanism with updated dark matter and dark energy.
S.Weinberg and P.Higgs found the mechanism that the $ SU(2)_W $ gauge fields $ W_{\mu}^a (x)(a=1,2,3)$ gain the energy by the sponteneous symmetry breaking of self-interacting potential and creat weak bosons $W_{\mu}^{\pm} (x) $ and $ Z_{\mu}^0(x).$
We consider that this mechnism is valid for all fermions and bosons.
We consider that fermion and  boson get the energy which corresponds to the change of stable point by the sponteneous symmetry breaking of the self-interacting potential formed by the  bosonic-state corresponding to each fied as proper mass  $(m_{pot}) $.

The  spectrum of proper mass $(m_{pot})$ of fermion obtained in this way has a fourth power structure. 
Since the electric charge of neutrino is zero,the self-iteracting potential space becomes very shallow.
This means that mass of neutrino is very small and close to the size of mass of dark matter.
It can be seen that the difference between mass spectrums of three generations of quark depends on the magnitude of two spinors and the inner product of two spinors(exactly spinor and anti-spinor).
 
\end{abstract}

\section{Introduction}
\quad This paper is presented for the following two reasons.
We want to solve the problems that were unknown in our previous paper.
We need to quickly overcome the aether thinking about mass.It was discovered that the electromagnetic interaction and the weak interaction are branches of the electroweak interaction.
Being able to understand two fundermental interaction in an unified manner was a breakthrough in the search for the laws of nature.
With the discovery of the electroweak interaction, the generation mechanism of the electromagnetic field $A_\mu$(x) and the mass creation mechanism of weak bosons $W_\mu^\pm$(x) and $Z_\mu^0$(x) were discovered.
This gave us the key to understanding the nature of electric charge and proper mass.The electromagnetic field $A_\mu$(x) and mass do not exist in the space-time region where the electroweak inrteraction holds.
Strictly speaking, a space-time region where the eletromagnetic field $A_\mu$(x) does not exist is a region where Minkowski space-time is not valid.

We consider the space-time where the electromagnetic field $A_\mu$(x) does not exist and Minkowski space-time is not valid as the quaternion space-time. 
Hamilton's quaternion is a number for consisting an one-dimensional real part and a three-dimensional imaginary part.
If there is no mass in space-time,spinor is two component Weyl spinor.
In quaternion space-time, the lepton field exists in one dimensional real space-time and the quark field exists in three dimensional imaginary space-time.
Bosonic-states composed of spinor and anti-spinor exist in real space-time.
We call the group of fermions and bosonic-states the Weyl spinor phase and define the Weyl spinor phase as dark energy.
If the quaternion space-time and the definition of dark energy are accepted,it is easy to understand that there are three generations of fermion,and that quark has three dimensional rotational symmetry and is confined,and that quarks exist in a mixture, and that the pressure of quark is negative.
 
There are two types of energy that can be transformed as proper mass.
One is our mechanism associated with the generation of the electromagnetic field $A_\mu$(x).
Although the amount of energy flow due to this mechanism is small,it is a mechanism for simulteneous generation of matter field,boson,dark matter,and dark energy.
 It is deeply related the composition and evolution of the universe.
Next, there is potential energy produced by the sponteneous symmetry breaking in the self-interacting potential space.
S.Weinberg and P.Higgs showed that the $SU(2)_W$ gauge fields can get this potential energy as proper mass.
We consider that this phenomenon is common to all field, and that this phenomenon occurs  at the same time.
The $SU(2)_W$ gauge fields gets energy through interacting with the Higgs doublet $H(x)=\displaystyle\left(\begin{array}{c}\Phi^+(x)\\\Phi^0(x)\end{array}\right)$. But, all fields directly get energy as proper mass $(m_{pot}).$
In section 2, we will describe about the quaternion space-time. 
In section 3, we will describe about the simultaneous generation of matter field,dark matter,dark energy, and boson, and about the early universe.
 In section 4,we will describe about  mass creation  of fermion.
In section 5, we will describe  discussion and conclusion.

\section{The quaternion space-time-The Weyl spinor phase-}
 \quad In the early universe,the $ SU(2)_W\times U(1)_Y$ symmetry holds,so the gauge fields $W_\mu^a(x)(a=1,2,3)$ and $B_\mu(x)$ exist.
This symmetrical universe change to an universe where massive weak bosons $W_\mu^\pm(x)$ and $Z_\mu^0(x)$ and the electromagnetic field $A_\mu(x)$ exist due to the sponteneous symmetry breaking caused by the Higgs doublet $ H(x)=\left(\begin{array}{c}\Phi^+(x)\\\Phi^0(x)\end{array}\right)$.
The sponteneous symmetry breaking phenomenon is that the stable point (vacuum) with the self-interacting potential space formed by the Higgs doublet $H(x)=\left(\begin{array}{c}\Phi^+(x)\\\Phi^0(x)\end{array}\right)$  change sponteneously 
and gauge fields obtain the potential energy corresponding to the amount of change as proper mass.
Strictly speaking,since mass and the electromagnetic field $A_\mu(x)$ do not exist,this is an area where the effectiveness of Minkowski space-time is questionable.
The problem of fermion mass generation is a problem in such a delicate space-time domain.
Therefore,we think about the space-time of the universe as follows.

\subsection {The quark field}
\quad We propose that the space-time in the region where mass and the electromagnetic field $A_\mu(x)$ do not exist is the quaternion space-time.
In this space-time ,"light velocity" is written as $c_{DE} $ and  "time" as $"t_{DE}"$.
The quaternion of Hamilton is a number consisting of an one-dimensional real part and a three -dimensional imaginary part.
In the quaternion space-time,lepton exist in the real space-time part and quark exist in a three-dimensional imaginary space-time part.
We propose this ideas about space-time based on the following four considerations.
The fact that fermions exist in three generations, is a reflection of the fact that three generations of quark exist orthogonally to each other in  three dimensional imaginary space-time.
The fact that quarks are essentially mixed,is a reflection of the fact that quark exists in three dimensional imaginary space-time that extends orthogonally to each other. 

The fact that quark has a color $SU(3)_C$ symmetry and is confined,is a reflection of the fact that quark exists rotationally symmetrically in three dimensional imaginary space-time,
and cannot be extracted into the real space.Furthermore, if quark is a complete fluid existing in three dimensional imaginary space-time,then the velocity of quark can be considered to be a pure imaginary number.
The relation between pressure $p$ and density $\rho$ of quark like this may match between pressure and density  predicted in dark energy.

Considering these arguments and the arguments in the next section,we consider two component Weyl spinor phase as dark energy.

Two component Weyl spinor exists in one-dimentional real space-time and three dimensional imaginary space-time,but has no electric charge and no mass,so it is a spinor whose  quantity
to characterize is unknown.
As discribed in the next section,when an electromagnetic field $A_\mu(x)$ appears and mass is generated, two Weyl spinors combine to generate a four component spinor.
We will treat the mixing of quark and proper mass creation approximately by four component spinor.

\subsection {Two component spinor and four component spinor}
\quad The fermion is combined into the family consisting of lepton and quark and there are three generations of the family.
These are 
$(\nu_e, e^-, u,d) $,$(\nu_\mu,\mu^- ,c,s )$ and $(\nu_\tau, \tau^-,t ,b)$.

These fields are characterized by electric charge and proper mass.
The equation of motion of four component spinor is described by the Dirac equation.
Since the structure of three generations is the same,we consider one family.
We write the four component spinor as follows.
$\nu_e=\left( \begin{array}{c}(\nu_e)_1\\ (\nu_e)_2\end{array}\right)$,$ e^-=\left( \begin{array}{c} (e^-)_1\\(e^-)_2\end{array} \right)$, 
$u=\left( \begin{array}{c} u_1\\ u_2 \end{array} \right)$ ,$d=\left( \begin{array}{c} d_1 \\ d_2 \end{array} \right)$.

Here, $(\nu_e)_i$, $(e^-)_i$, $u_i$ ,and $d_i$ $(i=1,2)$ represent two component spinor.
Each fermion has an anti-fermion
 $(\nu_e)^\ast$, $(e^-)^\ast$, $(u^\ast)$, $(d^\ast)$.
Bosonic-states consisting of fermion and anti-fermion are shown as following,
$\nu_e\cdot(\nu_e)^\ast$,$\nu_e\cdot(e^-)^\ast$,$e^-\cdot(\nu_e)^\ast$,$e^-\cdot(e^-)^\ast$,  
$u\cdot(u^\ast) $,$u\cdot(d^\ast)$,$ d\cdot(u^\ast)$,$d\cdot(d^\ast)$.  
There are eight sets of bosonic states.
The number of fermion(including anti-fermion) and bosonic-states in one family is both eight.

\subsection {Mixing of three families of fermion}
\quad The quaternion space-time is the space-time in which two component Weyl spinor exist.
A mixture of three generations of lepton exists in one-dimensional real part.
A mixture of three generations of quark exists in three dimensional imaginary part.
In this way,three generations of Weyl spinor exist in a mixture.
We make the following three approximations.
(1) Lepton and quark are considered as four component spinor. (2)The mixture of fermion belong to lepton and quark in the same family is approximated as the same. (3)We assume that three families
 exist orthogonally to each other.
We write the three families as $f_i=(l_i,q_i),(i,j=1,2,3)$.
Here, $l_i$ and $q_i$ represent the following
\[  l_1=\left( \begin{array}{c} \nu_e \\ e^-\end{array} \right), l_2=\left( \begin{array}{c} \nu_\mu \\ \mu^-\end{array} \right), l_3=\left( \begin{array}{c} \nu_\tau \\ \tau^-\end{array}  \right) \]
\[ q_1=\left( \begin{array}{c} u \\ d \end{array} \right),  q_2=\left( \begin{array}{c} c \\ s \end{array} \right),  q_3=\left( \begin{array}{c} t \\ b \end{array} \right). \]

When we express the mixed states of lepton $l_i$ and quark $q_i$ as $(l_i)^\prime$  and $(q_i)^\prime$ respectively,
mixture of lepton and quark can be writen as 

  \[ (l_i)^\prime=\sum_{j=1}^{3}V_{lij}l_j  \]
  \[ (q_i)^\prime=\sum_{j=1}^{3}V_{qij}q_j. \]
 \qquad\qquad\qquad\qquad where  $i=1,2,3$.\\
\quad Matrix($V_l$) represents a mixture of three generations of lepton in one-dimensional real space.
Since there is little mixing ,matrix($V_l$) is considered to be close to the identity matrix.
Matrix($V_q$) represents a mixture of three generations of quark in three dimensional imaginary space.
Here,this is consistent with the Cabibbo-Kobayashi-Maskawa matrix which has the same meaning.

\section{The creation mechanism of matter field, dark matter, and dark energy}
\quad The electric charge and proper mass are important physical quantities that characterize fields.
The discovery of the electroweak interaction and the Higgs mechanism clarified the process of generating these physical quantities.
It was revealed that the $SU(2)_W$ gauge fields $W_\mu^a(x)(a=1,2,3)$ would get the potential energy caused by the sponteneous symmetry breaking in the self-interacting potential space formed by the Higgs doublet $H(x)=\left( \begin{array}{c} \Phi^+(x) \\ \Phi^0(x) \end{array} \right) $,and become massive weak bosons $W_\mu^\pm(x)$ and $Z_\mu^0(x)$.
We showed that when the electromagnetic field $A_\mu(x)$ is generated, the local $U(1)_{em}$ gauge transformation becomes discontinuous and the energy flows into matter field, and proper mass is generated.
It was shown that matter field, dark matter and dark energy  are generated by this energy flow at the same time.
In this paper, we will discuss two types of the mass creation mechanism and generating components of the early universe.
 \subsection {The generation of the mass term and four component spinor}
\quad Let us consider the local $U(1)_{em}$ gauge transformation of the massless four component fermion when $A_\mu(x)$ appears.
Lagrangian density is

\[ L=\bar{\psi }(x)i\gamma ^\mu (\partial _\mu +iQA_\mu (x))\psi (x) .  \]

$Q$ is the electric charge of fermion.
 $\partial _\mu =\displaystyle{\frac{\partial }{\partial x_\mu }}\,$ $\,(\,\mu =0,1,2,3\,)$ and $\,\partial _t=\displaystyle{\frac{\partial }{\partial t}}\,$  are a differential to 4-vector $x_\mu$ and a differential to $t$.

Let us consider the local $U(1)_{em}$ gauge transformation for this Lagrangian density.
\begin{align*}
  \psi^\prime(x)&=e^{-i\alpha(x)}\psi(x) \\ 
  A^\prime_\mu(x)&=A_\mu(x)+\displaystyle\frac{1}{Q}\partial_\mu\alpha(x).
\end{align*}
Consider only the zeroth component
\begin{align*}
 \partial_0\psi^\prime(x)&=e^{-i\alpha(x)}( \partial_0\psi(x)-i(\partial_0\alpha(x))_K)\psi(x) \\
 A^\prime_0(x)&=A_0(x)+\displaystyle\frac{1}{Q}(\partial_0\alpha(x))_G.
\end{align*}
Here,$(\quad)_K$ and $(\quad)_G$  are the differentials of the kinetic term and the gauge field with respect to the phase $\alpha(x)$.
Lagrangian density is as follows,
\[ \bar{\psi}^\prime(x)i\gamma^0(\partial_0+iQA^\prime_0(x))\psi^\prime(x)=
\bar{\psi}(x)i\gamma^0(\partial_0+iQA_0(x))\psi(x)+\bar{\psi}(x)\gamma^0(\Delta\dot{\alpha})_{KG}\psi(x). \]
Here, \[ (\Delta\dot{\alpha})_{KG}\equiv(\partial_0\alpha(x))_K-(\partial_0\alpha(x))_G=(c_G-c_K)(constant)\dot{\alpha}.  \]
Here,$\dot{\alpha} \equiv \partial_t\alpha(x)$ and $c_K$ and $c_G$ represent the light velocity of the kinetic term side and the gauge transformation side,respectively.
If the energy $(\Delta\dot{\alpha})_{KG}$ flows in the direction $\gamma^0$=$\textstyle\left( \begin{array}{cc} 0&I \\ I&0 \end{array} \right)$ that connects two Weyl spinors,two Weyl spinors combine to form a four component spinor.
For this reason,we replace $(\Delta\dot{\alpha})_{KG}$ with $(\Delta\dot{\alpha})_{KG}\gamma^0$.
Thus,the second term becomes
  \[ (\Delta\dot{\alpha})_{KG}\bar{\psi}(x)\psi(x). \]
This mass term shows the existence of three types of fields as follows.\\
\quad (case a) If K=G, \\
  \qquad\qquad\qquad $c_K$=$c_G$=c or  $c_K$=$c_G$=$c_{DE}$.\\
Here, c is the light velocity and $c_{DE}$ is the velocity parameter in dark energy(the Weyl spinor phase).
\[ (\Delta\dot{\alpha})_{KG}=0.  \]
Since  there is no flow of energy,two Weyl spinors is massless and remain in dark energy. \\
\quad (case b) If $c_K$=$c_{DE}$ and $c_G$=c,\\
\[ (\Delta\dot{\alpha})_{KG}\equiv m_{A_\mu}   (\quad>0).\]
The fermion of proper mass $(m_{A_\mu}) $ is generated.
The mass term becomes
 \[ m_{A_\mu}\bar\psi(x)\psi(x). \] 
\quad (case c) If $c_K=c$ and $c_G$ =$c_{DE}$, \\
\[ (\Delta\dot{\alpha})_{KG}=-m_{A_\mu}.  \]
The mass term becomes
\[ -m_{A_\mu}\bar\psi(x)\psi(x). \]
\quad We defined this wrong sign mass term as dark matter.
Therefore,the same number of dark matter and fermion are generated.
The mass creation of boson can be discussed in the same way,but we will omit it.
Dark energy of boson is generated in half just like fermions.
Since the mass term of boson is square, dark matter of boson does not exist.
That is , dark matter is the fermion corresponding to the wrong sign mass term.
\subsection{ Problems of the early universe}
\quad Our mass creation mechanism is production mechanism of proper mass $(m_{A_\mu})$ by the generated energy flow with the appearance of the electromagnetic field $A_\mu(x)$.
But, our mechanism is the simultaneous generation of matter field,dark matter,dark energy,and boson.
So,it is deeply related various problems in the early universe.
 \subsubsection{ The composition of the universe}
\quad The ratio of the number of baryon,dark matter,and dark energy(the cosmological constant$(\Lambda)$) generated when the electromagnetic field$(A_\mu(x))$ appeared, can be calculated.
For example,the ratio would be 4 \% : 24\% : 72\% .

The counting method differs depending on whether baryon includes anti-baryon or not.
But, since the ratio of the number of generated fermion and dark matter is 1 to 6 by the definition of dark matter,even if the counting method changes, the results will be close to the composition ratio currently obtained.

\subsubsection{ Dark matter}
\quad Dark matter from our model can be written as below,
\[ -m_{A_\mu}\bar\psi(x)\psi(x).  \]
\quad As we will see in the next section,the size of $m_{A_\mu} $ is smaller than mass of neutrino.
There are six times as many as baryons in the universe.
Since the sign of mass of matter field and dark matter are opposite,a gravitational repulsive force acts between matter field and dark matter.
There are three possibilities regarding to realm of dark matter existence. \\
\quad (a)It is expected that the large-scale structure of the universe was gradually formed by the gravitational repulsive force acting between matter field and dark matter.\\
\quad (b)It is expected that there are limited areas where dark matters of relatively low energy and high number density are surrounded by matter field.\\
\quad (c)It is expected that there are high energy dark matter that coexists with matter fields.
Since the gravitational repulsive force is very weak,dark matter will coexist as neutrino which does not act on weak interaction
\subsubsection {Dark energy (The cosmological constant $(\Lambda)$)}
\quad In the mechanism of our mass creation, the electromagnetic field$(A_\mu(x))$ gerneration process is a process in which the cosmological constant($\Lambda$) decrease by half.
According to general relativity,for the universe consisting only of uniform and isotropic dark energy (the cosmological constant($\Lambda$)), it is 
known that the size of the universe changes according to the Friedmann equation.
The size of the universe is expanding exponentially.

\subsubsection {Boson }
\quad According to our mechanism,bosons with mass similous to quarks are generated about the same number of dark matter.  
It is interesting to study the possibility of black hole formation by bosons.
\section{ The mass creation mechanism --The sponteneous symmetry breaking in the self-interacting potential space}
\quad S.Weinberg and P.Higgs showed that the mass creation of the $SU(2)_W$ gauge fields $W_\mu^a(x)$\quad$(a=1,2,3)$ can be explained by the spontaneous symmetry breaking in the self-interacting potential space.
Energy is generated as a difference of change of stable point(vacuum) in the self-interacting potential space formed by the Higgs doublet$ H(x)=\displaystyle\left(\begin{array}{c}\Phi^+(x)\\\Phi^0(x)\end{array}\right)$.
The gauge fields $W_\mu^a$(x)\quad$(a=1,2,3)$ get this energy through interacting with the Higgs doublet $H(x)=\displaystyle\left(\begin{array}{c}\Phi^+(x)\\\Phi^0(x)\end{array}\right)$, and massive weak bosons  $W_\mu^\pm$(x) and  $Z_\mu^0$(x) are generated.
We consider that not only the gauge field but also felmion field and boson field gain proper mass through the sponteneous symmetry breaking phenomenon.
We consider that fermion and boson get energy which corresponds to the change of the stable point(vacuum) in the self-interacting potential space formed by bosonic-state corresponding to each field as proper mass $(m_{pot})$. 
In the end,proper mass becomes $(m_{A_\mu}+m_{pot} )$,so the Lagrangian density of fermion can be written as follows.
 \[  L=\bar{\psi}(x)(i\gamma^\mu\partial_\mu+m_{A_\mu}+m_{pot})\psi(x). \]
  \qquad Where $ m_{pot} \equiv V_{self}. $

\subsection{The proper mass $m_{pot}$}
\quad When the symmetry is broken sponteneously in  the self-intertacting potential space formed by the Higgs doublet $H(x)=\displaystyle\left(\begin{array}{c}\Phi^+(x)\\\Phi^0(x)\end{array}\right)$
 ,we assume that every field has a corresponding self-interacting potential space and the symmetry in the space is broken sponteneously at the same time.
The self-interacting potential of bosonic state $\Phi$ is 

 \[   V_{self}(\Phi(x))=\mu^2\Phi(x)\Phi^*(x)+\lambda(\Phi(x)\Phi^*(x))^2 .  \]
 Where $\lambda>0$ and $\mu^2>0$. 

When the symmetry is broken $\lambda>0$ and $\mu^2<0$ ,

  \[    V_{self}(\Phi(x))=\lambda(|\Phi(x)|^2-\frac{1}{2}|\Phi_0|^2)^2-\frac{1}{4}\lambda|\Phi_0|^4.   \]
 Where $ |\Phi_0|^2$=$-\frac{\mu^2}{\lambda}$.

It is thought that the symmetry has changed from the stable point(vacuum) of ($\lambda>0$,$\mu^2>0$) to the stable point(vacuum)of ($\lambda>0$,$\mu^2<0$) sponteneously.
Field get this energy $\frac{1}{4}\lambda|\Phi_0|^4$ as proper mass $(m_{pot})$.
Therefor, the total proper mass of the field becomes $(m_{A_\mu}$+$m_{pot})$.
As we will see later ,the size of $m_{A_\mu}$ is small. 
And, $m_{pot}$ depends on $|\Phi^0|$ to the fouth power.
Therefore, the proper mass $(m_{pot})$ is characterized by a fourth power structure.
 
\subsection{The mass creation of fermion}
\quad We consider the mass creation of fermions.
There are three generations,but the structure is the same. So,we consider the first generation $(\nu_e, e^-, u, d)$.
The proper mass and the electric charge of field are closely related. Therefore,the electric charge of field and that of corresponding bosonic-state must be the same.

\subsubsection{The bosonic-states of leptons: $\Phi^0$ and $\Phi^-$}
\quad Bosonic-states of leptons ($\nu_e$ and $e^-$) are composed from spinors ($\nu_e$ and $e^-$) and their anti-spinors ($(\nu_e)^*$ and $(e^-)^*$) and become as follows.\\
 The bosonic-state corresponding to neutrino is $\Phi^0$=$(|\nu_e|^2-|e^-|^2)$.\\
 The bosonic-state corresponding to electron is $\Phi^-$=$(e^-\cdot(\nu_e)^*)$.\\
Therefore,$m_{pot}$ of  leptons ($\nu_e$ and $e^-$) is as follows.
 \begin{align*}
  m_{pot}(\nu_e)=\frac{1}{4}\lambda|\Phi_0^0|^4  \\
  m_{pot}(e^-)=\frac{1}{4}\lambda|\Phi_0^-|^4.  
\end{align*}
where $ |\Phi_0^0|^2=(|\nu_e|^2_0-|e^-|^2_0)^2 $ and $    |\Phi_0^-|^2=|(e^-\cdot(\nu_e)^*)_0|^2. $

The measured values of mass of leptons with its fourth power structure are as follows.

 \begin{align*}
  m_e&=0.51  MeV        &=(0.845)^4 MeV    \hspace{1cm}      m_{\nu_e}    &< 2 eV         &=(0.0376)^4 MeV  \\
  m_\mu&= 105.65 MeV &=(3.206)^4 MeV    \hspace{1cm}     m_{\nu_\mu} &<0.19 MeV   &=(0.6602)^4 MeV \\ 
 m_\tau&=1777 MeV    &=(6.4926)^4 MeV  \hspace{1cm}     m_{\nu_\tau} &<18 MeV     &=(2.0598)^4 MeV  
\end{align*}

Since the electric charge of neutrino is zero,the self-interacting potential space has an unique structure and becomes shallow.

If $|e^-|^2_0\approx|\nu_e|^2_0$, $m_{\nu_e}\approx m_{A_\mu}$.
If mass of neutrino is small, absolute value of mass of dark matter is also small.
 
\subsubsection{The bosonic-states of quark: $\Phi^{\frac{2}{3}}$ and $\Phi^{-\frac{1}{3}}$}
\quad Since electric charges of quark($u$ and $d$) are $\frac{2}{3}e$ and $-\frac{1}{3}e$, electric charges of corresponding bosonic-states ($\Phi^{\frac{2}{3}}$ and $\Phi^{-\frac{1}{3}})$ must be the same.
But, electric charges of bosonic-states $\Phi^+$=$(u\cdot(d^*))$ and $\Phi^-$=$(d\cdot(u^*))$ are +e and -e, so bosonic-states ($\Phi^{\frac{2}{3}}$ and $\Phi^{-\frac{1}{3}})$ of quark(u and d) are constructed by transforming bosonic-states ($\Phi^+$ and $\Phi^-$). \\
We consider the following transformation from $(\Phi^+;\Phi^-)$ to $(\Phi^{\frac{2}{3}}.\Phi^{-\frac{1}{3}}).$

\[  \left( \begin{array}{c} \Phi^{\frac{2}{3}} \\ \Phi^{-\frac{1}{3}} \end{array} \right) = \begin{bmatrix} a&b \\ c&d \end{bmatrix} \left( \begin{array}{c} \Phi^+ \\ \Phi^- \end{array} \right) \]
Here, four constants(a,b,c,d) are the parameters to be determined.
We find four parameters of matrix using the following two conditions.
The first condition is the conversion from electric charges $(1,-1)e$ to electric charges $(\frac{2}{3},-\frac{1}{3})e$.
The second condition is that this transformation is a rotation transformation.
       \[     \begin{bmatrix} a&b \\ c&d \end{bmatrix} = \begin{bmatrix} A\cos\theta&B\sin\theta \\ -B\sin\theta&A\cos\theta \end{bmatrix} \]
The transformation that satisfies two conditions is as follows.

\[ \left( \begin{array}{c} \Phi^{\frac{2}{3}} \\ \Phi^{-\frac{1}{3}} \end{array} \right)= \begin{bmatrix} \frac{1}{2}&-\frac{1}{6} \\ \frac{1}{6}&\frac{1}{2} \end{bmatrix}  \left( \begin{array}{c} \Phi^+ \\ \Phi^-\end{array} \right) \]

Here,since $\Phi^+$=$(u\cdot(d^*))$ and $\Phi^-$=$(d\cdot(u^*))$,the relationship $\Phi^+$=$(\Phi^-)^*$ and $\Phi^-$ =$(\Phi^+)^*$ holds true.
If we write $\Phi^+$ as $\Phi^+$=$Re\Phi^+$+i$Im\Phi^+$,
$\Phi^{\frac{2}{3}}$  and $\Phi^{-\frac{1}{3}}$ can be expressed  using these $Re\Phi^+$ and $Im\Phi^+$ as follows.
\[
 \Phi^{\frac{2}{3}}=\frac{1}{3}(Re\Phi^+)+i\frac{2}{3}(Im\Phi^+).   
\]
\[
\Phi^{-\frac{1}{3}}=\frac{2}{3}(Re\Phi^+)-i\frac{1}{3}(Im\Phi^+).
\]

From the above results,the proper mass of quark (u and d) is as follows.
 \[  
m_{pot}(u)=\frac{1}{4}\lambda|\Phi_0^{\frac{2}{3}}|^4=\frac{1}{4}\lambda(\frac{1}{9}(Re\Phi_0^+)^2+\frac{4}{9}(Im\Phi_0^+)^2)^2.  
\]
\[ 
m_{pot}(d)=\frac{1}{4}\lambda|\Phi_0^{-\frac{1}{3}}|^4=\frac{1}{4}\lambda(\frac{4}{9}(Re\Phi_0^+)^2+\frac{1}{9}(Im\Phi_0^+)^2)^2
\]
The measured values of mass of quarks with its fouth power structure are as follows. 

 \begin{align*}
 m(u)&=3MeV&=(1.3)^4MeV \hspace{1cm}  m(d)&=5MeV&=(1.5)^4MeV  \\
 m(c)&=1.3GeV&=(6.0)^4MeV \hspace{1cm}  m(s)&=100MeV&=(3.16)^4MeV \\
 m(t)&=172GeV&=(20.3)^4MeV \hspace{1cm}  m(b)&=4GeV&=(8.0)^4 MeV 
\end{align*}

Although two bosonic-states of quark (u and d) are similar,but $m(u)<m(d)$ in the first generation, $m(c)>m(s)$ in the second generation and $m(t)>m(b)$ in the third generation.
The reason why the appearance of mass spectrum looks different among these three generations can be explained by the difference between the two $m_{pot}$ of same family.

 \[  \Delta(u-d)\equiv m_{pot}(u)-m_{pot}(d)   \\
                  =\frac{5}{108}\lambda|\Phi_0^+|^2(-(Re\Phi_0^+)^2+(Im\Phi_0^+)^2)  \]

In the first generation,the real part$(Re\Phi_0^+)$ of $\Phi_0^+$=$(u\cdot(d^*))_0$ is greater than the imaginary part $(Im\Phi_0^+)$.
In the second and third generation,the imaginary part $(Im\Phi_0^+)$ of $\Phi_0^+$=$(u\cdot(d^*))_0$ is greater than the real part $(Re\Phi_0^+)$.
It can be seen that the mass spectrum of quark  depends on the inner product of ($u$ and $d^*$) as well as the size of spinors $(u$ and $d^*)$.

\section{Conclusion and discussion}
\quad The electroweak interaction is widely accepted as an unified theory of the electromagnetic interaction and the weak interaction.
The appearance of $A_\mu(x)$ suggests the existence of a space-time other than Minkowski space-time.
 We considered the unknown space-time to be the quaternion space-time (Hamilton space-time) including three dimensional imaginary space-time.
The three generations of quark are orthogonal to each other and exist mixed together in three dimentional imaginary space-time.
This picture of quark can explain  the confinement of quark and the mixing of quark and the existence of three generation.
And it can be inferred that the kinetic energy of quark in this space-time is negative.\\
\quad In the quaternion space-time,lepton exists in one-dimensional real space-time and quark exists in three dimensional imaginary space-time.
The bosonic-state consisting of spinor and anti-spinor exists in one-dimensional real space-time.
We defined the Weyl spinor phase as dark energy.
 The electromagnetic field $A_\mu(x)$ is generated i.e. the light velocity becomes discontinuous.
This means that a flow of energy is generated according to our mass creation mechanism.
Through this energy flow,two Weyl spinors are combined and a massive four component spinor is generated.
At this time,matter field,dark matter,dark energy,and boson are generated at the same time.
The ratio of the number of generated matter field,dark matter,and dark energy can be calculated as 4\%:24\%:72\%.
This value is close to the observed value.\\
\quad In the finding of the electroweak interaction,it was shown that the gauge fields $W_\mu^a(x) (a=1,2,3)$ obtains the energy caused by the sponteneous symmetry breaking in the self-interacting potential space formed  by the Higgs doublet $H(x)=\displaystyle\left(\begin{array}{c}\Phi^+(x)\\\Phi^0(x)\end{array}\right)$ as  mass.
We thought that a bosonic-state with the same electric charge as field exists,and in the self-interacting potential space formed by the corresponding bosonic-state,this phenomenon occured not only in the gauge field but also in all fields(fermion and boson) at the same time.
The proper mass$(m_{pot})$of potential energy generated by this phenomenon has the fourth power structure.
Since the electric charge of neutrino is zero,the self-interacting potential space becomes a shallow structure.
Therefore,proper mass of neutrino becomes very small.
And it is not much different from the size of mass of dark matter.
The spectrum of three generations  of quark depends on the difference in size of spinors between generations and the difference in inner products of two spinors(exactly spinor and anti-spinor) in the same family between generations.
There are several important problems to be solve in the future.\\
(1)dark energy\\
\quad It is important to find the law of motion for Weyl spinor in the the quaternion space-time  to calculate the mass of fermion and the kinetic energy of quark.\\
\quad And it is nececcary to clarify the relationship between the Higgs doublet $H(x)=\displaystyle\left(\begin{array}{c}\Phi^+(x)\\\Phi^0(x)\end{array}\right)$  and the Weyl spinor phase.\\
(2)dark matter\\
\quad It is interesting to investigate whether a star-like object is formed in a region where only dark matter exists.\\
(3)boson\\
\quad Since a large amount of boson is generated in our mass creation mechanism,it is important to study the formation of black hole by boson.\\

\end{document}